# Formation Flying Techniques for the Virtual Telescope for X-Ray Observations

**Rankin Kyle[a], Shah Neerav[b], Krizmanic John[c], Stochaj Steven[d], Naseri Asal[e]**

[a] Department of Mechanical and Aerospace Engineering, New Mexico State University, krankii@nmsu.edu
[b] Goddard Space Flight Center, NASA neerav.shah-1@nasa.gov
[c] CRESST, NASA, GSFC, University of Maryland Baltimore County, John.F.Krizmanic@nasa.gov
[d] Klipsch School of Electrical and Computer Engineering, New Mexico State University, sstochaj@nmsu.edu
[e] Program Manage/Systems Engineer, Space Dynamics Laboratory, asal.naseri@sdl.usu.edu

## Abstract

The Virtual Telescope for X-Ray Observations (VTXO) is an Astrophysics SmallSat mission being developed to demonstrate 10-milliarcsecond X-ray imaging using a Phase Fresnel Lense (PFL)-based space telescope. PFLs promise to provide several orders of magnitude improvement in angular resolution over current state of the art X-ray optics. However, PFLs for astronomical applications requires a very long focal length, for VTXO the focal length is estimated to be in the range of 0.5 km to 4 km. Since these focal lengths are not feasible on a single spacecraft, the proposed solution is to use two separate spacecraft, one with the lense(s), and the second with an X-ray camera. These two spacecrafts will then fly in a formation approximating a single rigid telescope. In order to achieve this configuration, the two spacecraft must maintain the formation a focal length distance apart, with centimeter level control, and sub-millimeter level knowledge requirements. Additionally, for the telescope to achieve sufficient exposure time, the system must keep the telescope axis pointed at a fixed target on the celestial sphere for durations on the order of a few hours. VTXO's system architecture calls for two CubeSats to operate in a highly eccentric Earth orbit with one of the spacecraft's traveling on a natural keplarian orbit. The second spacecraft will then fly on a "pseudo" orbit maintaining a fixed offset relative to the celestial sphere during observations. Observations with this system will occur near apogee where differential forces on the spacecrafts are minimal which in turn minimizes fuel consumption. The "pseudo" orbit where the second spacecraft is located is not quite a stable orbit requiring small continuous propulsion to maintain formation. This paper and presentation will overview VTXO's system architecture, additionally it will look in depth at the formation flying techniques, including fuel consumption, and methods for establishing, and maintaining the formation. Beyond its use in X-ray astronomy, these formations flying techniques should eventually contribute to the development of distributed aperture telescopes, with imaging performance orders of magnitude better than the current state of the art.

## Acronyms/Abbreviations

VTXO – Virtual Telescope for X-ray Observations
PFL – Phase Fresnel Lense
MMS – Magnetosphere MultiScale
TRL – Technology Readiness Level
GTO – Geostationary Transfer Orbit

## 1. Introduction

VTXO (Virtual Telescope for X-ray Observations) is flying a new form of X-ray optic known as the Phase Fresnel Lense (PFL). This lense which works on the basis of diffraction, should be capable of achieving near diffraction limited imaging at a specific X-ray energy. Based on VTXO's preliminary design with an ~2.5cm lens, it should be capable of improving on the angular resolution of Chandra, the current state of the art telescope, by a factor of 10. However, PFLs have a large focal length in excess of half a kilometer for VTXO. Since it is not practical to build a spacecraft that long, The VTXO mission will fly two spacecraft, the first with at least one PFL, and the second with an X-ray camera flying in a formation approximating a rigid telescope structure.

VTXO will first demonstrate the ability to create a useful X-ray telescope utilizing a PFL. Second, VTXO should be the highest resolution X-ray telescope in existence at the times of its launch and preform high angular resolution measurements on some of the brighter X-ray sources in the sky. Finally, VTXO will demonstrate the formation flying techniques required to build the next generation of very large distributed component telescopes.

### 1. Phase Fresnel Lenses

The major challenges in building X-ray optics are the result of the very short X-ray wavelength. This results in strict manufacturing tolerances for the optics to focus X-rays precisely enough. Additionally, at these wavelengths, most known materials have to efficiently focus incident X-rays (at a specific energy) into a primary focus [3, 4]. Laboratory tests of MEMS-fabricated PFLs have demonstrated near diffraction-limited imaging (~20 milli-arcseconds FWHM) in the X-ray band with efficiencies ~30% or better [5] [6] [7] [8]Figure 1 shows an SEM of one of the tested PFLs. Initial development on PFL-based acromats have demonstrated 10's of milli-arcsecond angular resolution over a much wider X-ray energy range than simple PFLs [7]. While PFL development will continue, the current state is that an X-ray PFL of several cm in size with at least 10 milli-arcsecond angular resolution and high transmission efficiency is straightforward to fabricate. However, one feature of PFLs is that the angular resolution of the lens improves with an increased focal length. In practice VTXO's lens will require a focal length of at least 0.5 km (to achieve 10's of milli-arcsecond angular resolution) and a focal length of 4 km would provide significantly improved science return (achieving even better angular resolution).

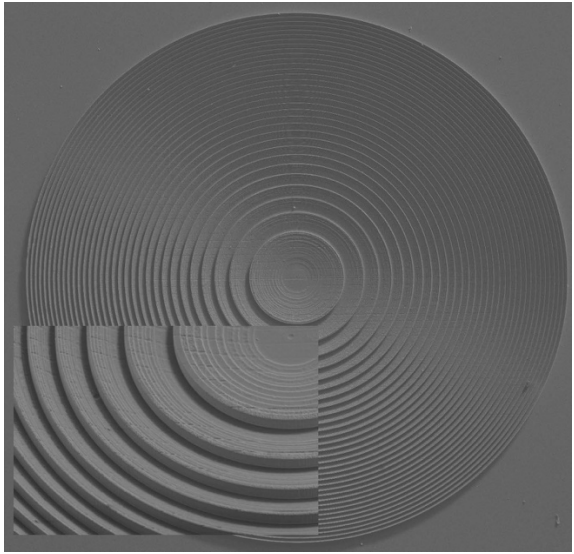

*Figure 1: SEM image of a prototype PFL [16].*

indexes of refraction very near 1.0, this means that if one used a traditional mirror like those in a visible wavelength telescope the x-ray would pass through the mirror, rather than reflecting. In order to overcome this, current generation X-ray telescopes such as Chandra [1] utilize nested grazing incidence mirrors oriented at a very low incidence angle to reflect the

## 2. VTXO Concept of Operation

The Concept of operations for the observation segment of the VTXO mission as described in Figure 2 is for two spacecraft, the Detector Satellite carrying the X-Ray camera, and the Optics Satellite carrying incoming X-rays at small angles to focus the light. This requires precise alignment of the complex mirror structure, and the need to polish these mirrors to smoothness of an angstrom level [2]. It is not currently possible to build these mirrors with adequate precision to even approach the diffraction limit with these X-ray optics.

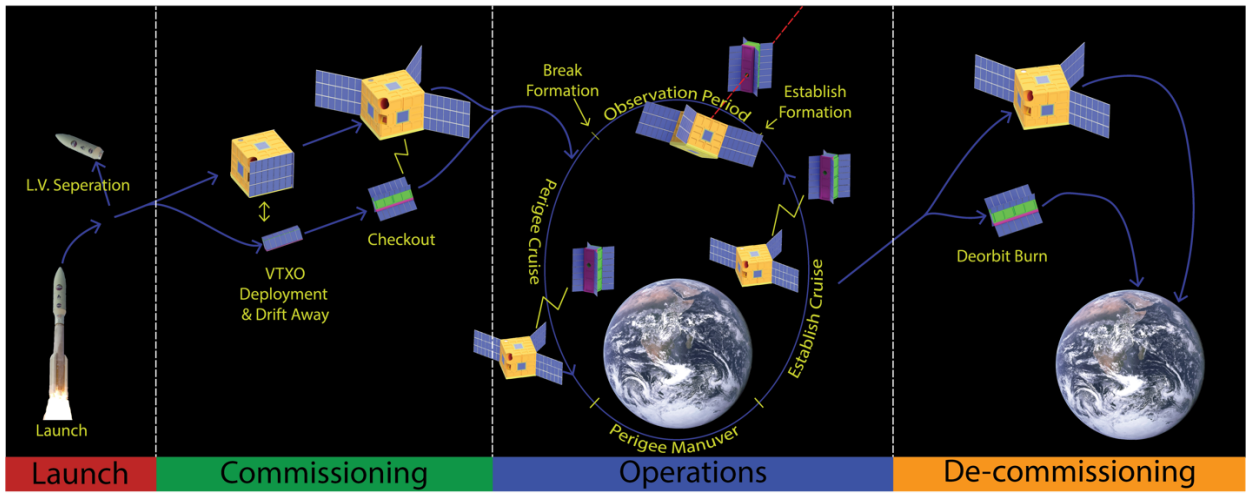

*Figure 2: VTXO Concept of Operations*

The PFLs used on VTXO offer significant improvement on the possible angular resolution compared to other X-ray optics. PFLs use diffraction the X-Ray lense(s) to fly in formation at least 0.5 km apart, with a precise alignment between the two spacecraft to make telescopic observations possible. This observation period will occur for a few hours while the spacecraft are near the apogee of a highly elliptical orbit where the gravity gradient is minimal. Then, the two spacecraft will break formation and perform a series of maneuvers while the two-spacecraft travel through perigee to re-align the two spacecraft for the next observation period out near apogee.

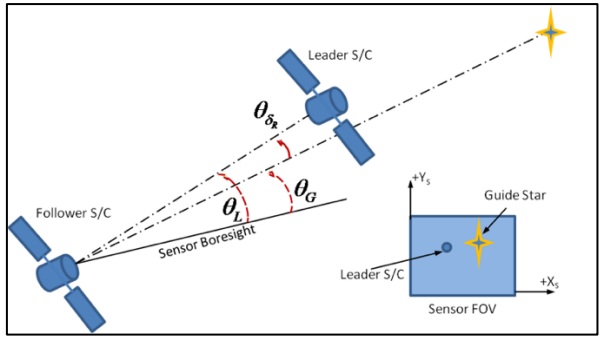

*Figure 3: Relative Navigation Scheme [12]*

### 3. Navigation Sensors

VTXO utilizes three types of sensors to meet the navigation requirements. For precision alignment, a relative navigation sensor based on a star tracker is used for precision alignment, as shown in Figure 3. This sensor is located on the Detector Satellite, and images a set of beacons on the Optics Satellite and compares them to the location of the background star field. This provides the absolute pointing direction of the telescope. VTXO baselines the NISTEx II star tracker, which has a designed angular resolution of ~60 mili-arc seconds. A separation distance of 0.5 km corresponds to a relative position knowledge normal to the telescope axis of ±145 microns.

Ranging during operations is performed by radio utilizing an intersatellite link to determine range between the spacecraft. This system should be able to operate at accuracies on the order of a few 10's of centimeters [9]. This easily meets the requirements for VTXO. There may be an alternative option to use onboard GPS for range measurement, however, GPS accuracy when above the GPS constellation such as will occur during observations may be borderline acceptable and further study would be required to determine if it will fully meet VTXO's ranging requirements.

Finally, for navigating during the drift around segment, the primary accuracy requirement is that the system must provide sufficient accuracy for the fine guidance sensor on the Detector Sat to acquire the Optics Sat. Based on the NISTEX II's 5 deg field of view, at 0.5 km separation, the rough alignment system must achieve a relative navigation solution better than ±43 m. In order to achieve this, VTXO will use a GPS on each satellite along with an inter satellite cross link to determine the relative positions of the two vehicles. GPS is typically capable of achieving position solutions on the order of 1 m of accuracy, which easily exceeds the 43 m requirement. There are some concerns that near VTXO's apogee of 42,164km (assuming GTO) is above the GPS constellation of ~20,180km. However, it has recently been demonstrated by the Magnetospheric Multiscale (MMS) mission that GPS fixes with better than ±15 m accuracy can be obtained up to, and above the altitude of VTXO's perigee [10]. Even if GPS acquisition at apogee proves problematic, by combining GPS measurements obtained at lower altitude, along with the ranging radio, and an inertial navigation system similar to what is used on MMS, an adequate navigation should be obtainable.

### 4. Propulsion Systems

VTXO requires a low variable thrust propulsion system. This can be achieved ether by using small throttleable systems, or by rapidly pulsing a fixed thrust engine. A high TRL (Technology Readiness Level) cold gas system based on the thruster currently in flight on JPL's MarCo [11] has been baselined for this mission. There is also the possibility of using either electrospray thrusters, or a mono-prop system, these could provide significant advantages in terms of mission life but come with the risk of flying lower TRL systems.

### 5. Control Systems

It has been determined that VTXO will be operating in an approximately linear control environment [12] [13]. Linear control systems are well understood and poses little risk. Assuming that the navigation sensors can provide sufficient accuracy, and the propulsion system can provide a sufficient thrust range, the control system is a well understood problem, which can be readily solved utilizing traditional linear control systems such as a PID controller [12].

### 6. Flight Dynamics

Flight Dynamics has been a major area of study in developing this mission concept. In order to maintain the rigid formation with a fixed pointing direction necessary for observations, one spacecraft can travel on a natural orbit trajectory, while the second spacecraft is inherently not on a natural trajectory. This results in a small gravity gradient between the two spacecraft, which the detector sat must compensate for by providing low level continuous propulsion. In order to minimize fuel consumption, the desired orbit is one where the gravity gradient can be minimized. As can be seen in Figure 4 in Earth orbit, due to the reduced gravity gradient, the fuel consumption required to maintain the formation will drop off by approximately one over the cube of the distance between the Earth and the Optics Sat.

Given the inherent limitations of rideshare launches, it was determined that the best orbit for the VTXO baseline is a Geostationary Transfer Orbit (GTO). This was selected, as it is the highest altitude launch available, that still has a high frequency of launch availability.

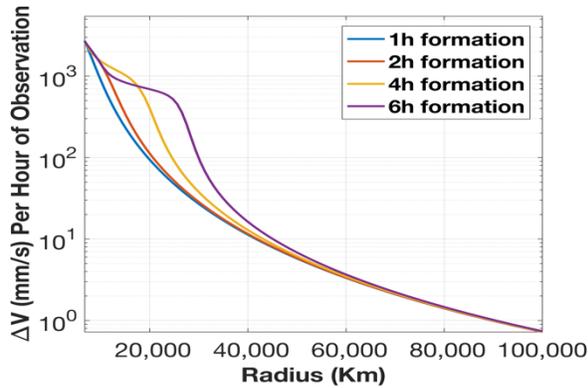

*Figure 4: Plot showing the fuel consumption required to maintain formation at different elevations above the Earth.*

## 7. Estimate of Formation Time per Orbit

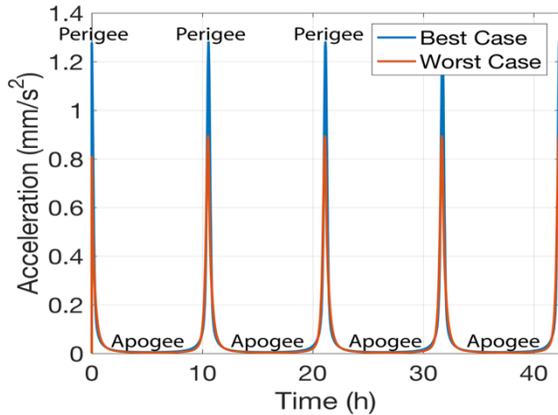

*Figure 5: Acceleration generated by the prop system over four orbits.*

An analysis of fuel consumption required to maintain the observation formation was performed along the entirety of a GTO orbit. Figure 5 shows the acceleration generated by the propulsion system. It is very clear that holding formation near perigee results in fuel consumptions dramatically higher then holding formation near apogee.

By integrating under the acceleration curve with respect to time, over an integral centered on apogee, it becomes possible to calculate the required Delta V of the system. Figure 6 shows the amount of fuel being consumed by holding the formation for various times and pointing directions. All of these formations are centered on the Optics Sat's apogee. This plot has been normalized to Delta V per hour of observation. As would be expected the longer the formation is held, the larger the penalty in terms of fuel consumption becomes. However as shown in Figure 6, the penalty remains modest for up to several hours of observation. Meaning formations of over 4h are reasonable.

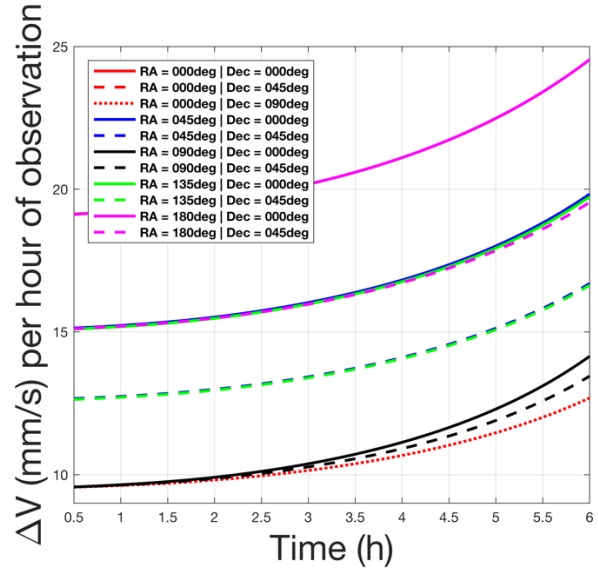

*Figure 6: Delta V per hour of observation as a function of the observation time and pointing direction.*

The second half of the fuel consumption problem is looking at the fuel consumption required to re-establish the observation formation during the coast through perigee. In order to understand this problem, a simulation was performed with random formation times, initial pointing, and final pointing directions. The results of these simulations are shown in Figure 7, and generally are a function of the angle between the two pointing directions, and the amount of time the formation is held for. Interestingly, as the formation time increases, the repoint Delta V tends to go down,

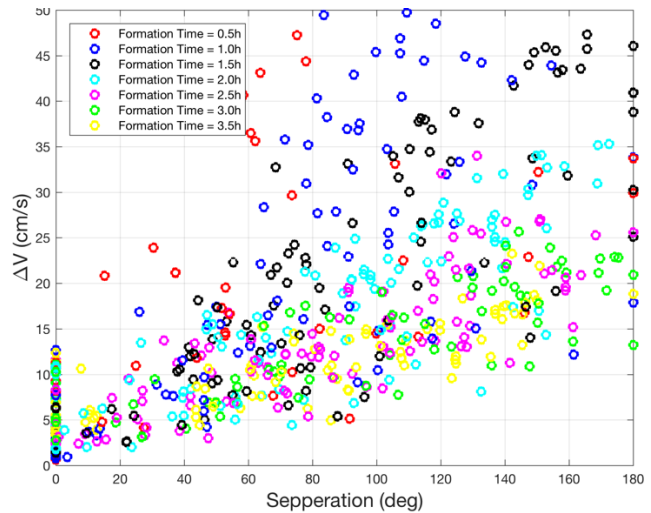

*Figure 7: Fuel Consumption per orbit to re-point Telescope relative to the angular distance between the pointing targets.*

resulting in an opportunity to increase the observation period.

As can in Figure 7, there are a great deal of variables that affect the overall fuel consumption, and the length of time that a formation can be held. However, it seems reasonable to conservatively assume that on average observations can be performed for around four hours each science orbit, and each science orbit would consume 20cm/s of Delta V. With careful planning of the observation sequence, better numbers may be achievable.

## 8. Target of Opportunity Analysis

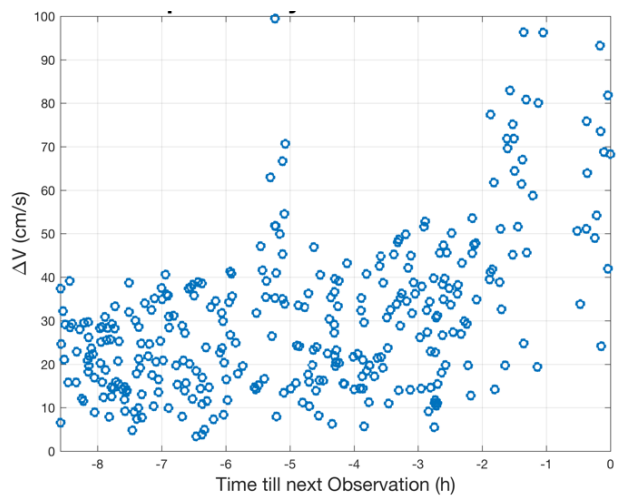

*Figure 8: Point Solution showing total fuel consumption for orbit where a repoint occurs.*

In order to estimate how quickly the telescope could be repointed at target of opportunity, such as a transient X-ray source in the sky, a simulation was run, where during a standard repoint sequence as described above, where a 2h observation is performed on either side of the repoint sequence, and the observations are made during those windows with two random pointing directions. Then at various points in the sequence, a command is made to repoint VTXO to a new pointing direction. As seen in Figure 8, only a modest increase in fuel consumption on the order of 2 – 3 times a typical orbit is seen for a repoint up until approximately 2h before the new formation is established. After the -2h mark, the fuel consumption rapidly increases, and goes off the scale of this plot, with some solutions as high as 1,500 cm/s being seen. It is worth noting that the peak just prior to the -5h mark is a result of the simulations repoint maneuver, and the simulation's perigee maneuver being placed into very close proximity of each other. In practice this should be avoidable by improving the maneuver timing.

Given that it is typically possible to repoint the VTXO till about two hours before the beginning of the formation, in most cases a repoint to a target of opportunity should be achievable at the worst case 12h after the command is sent. The worst case would occur if the command is uploaded right after VTXO enters the region where it is no longer possible to repoint for the upcoming formation.

## 9. Estimate of Mission Lifetime

VTXO's lifetime has a hard limit set by available propellent. The mission is divided into three primary segments commissioning, operations, and de-commissioning. The commissioning segment will consist of both spacecraft performing a maneuver to raise the perigee above the bulk of Earth's atmosphere, and then a sequence of maneuvers to correct for any drift away introduced by the launch vehicle separation. This initial sequence has the most unknowns, as it is highly dependent on the deployment conditions. The second segment is the operations phase, which has been assessed in previous sections. Finally, during the De-Commissioning phase, a maneuver performed by both spacecraft to lower their respective perigees far enough into earth's atmosphere to cause both spacecraft to deorbit in a relatively small amount of time.

*Table 1: Delta V Budget*

| Maneuver | Optics Sat | Detector Sat |
|---|---|---|
| Perigee Raise: | 10m/s | 10m/s |
| Establish Initial Formation[1]: | 10m/s | 10m/s |
| Operations: | 0 | 65m/s |
| De-Orbit: | 10m/s | 10m/s |
| **TOTAL:** | 30m/s | 95m/s |

*Table 2: Mission Timeline*

| Mission Element | Time Consumed | Total Time Elapsed | Observation Time |
|---|---|---|---|
| Launch / Deployment | 1 day | L +1 day | |
| Commissioning | 60 days | L +61 days | |
| Operations | 110 days | L +171 days | 747h |
| De-Commissioning | 5 days | L +176 days | |

Note that the mission lifetime is based on the baselined cold gas propulsion system. Based on a preliminary

---

[1] Values are rough estimates and are highly dependent on initial deployment conditions, and time required for the commissioning phase

estimate a mono-prop system could extend the observation section to around 830 days. An electrospray system could extend the fuel availability to nearly 16 years point at which point propellent is no longer the limiting factor in mission lifetime.

## 10. Alternative Orbits

A GTO orbit was selected based on availability of launches but, is not the only option for VTXO. If a launch can be obtained, higher Earth orbits such as a Molynia, Tundra, or Lunar Transfer Orbit are all viable, and would reduce fuel consumption, along with the corresponding increase in mission life, while also reducing trapped-particle radiation concerns. Additionally, ether an Earth drift away orbit or a liberation point orbit would be an excellent option for formation flying reducing fuel consumption by more than an order of magnitude, and they would reduce VTXO's radiation dose relative to a GTO. However, these orbits may come with a significant increase of complexity and cost, particularly in terms of communications systems, potentially requiring the use of the Deep Space Network.

## 11. Baseline CAD Model

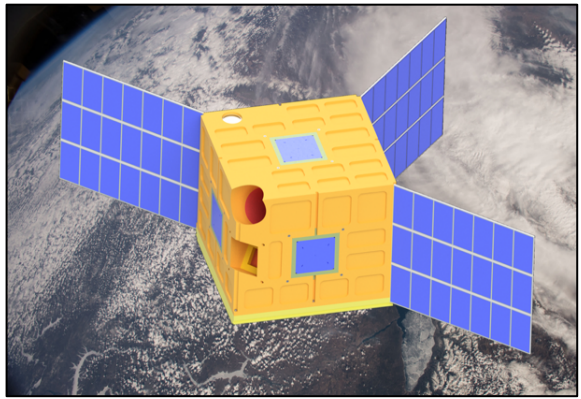

*Figure 9: Detector Sat Flying over the Earth*

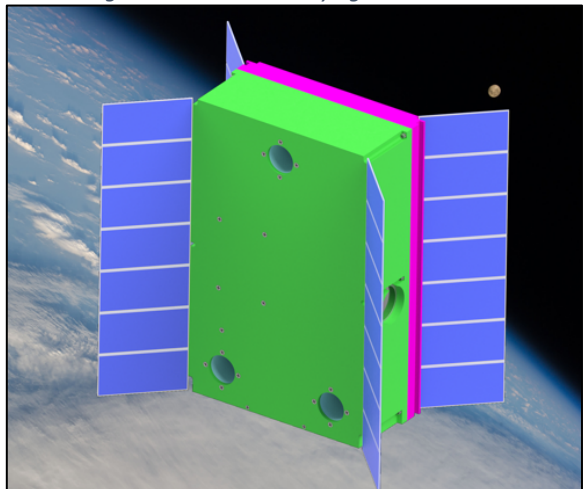

*Figure 10: Detector Sat Flying over the Earth*

The baseline design for the VTXO spacecraft includes two vehicles, one 6U CubeSat, and a second 27U CubeSat. Both are designed to the specifications for the Planetary Systems Corporation deployers.

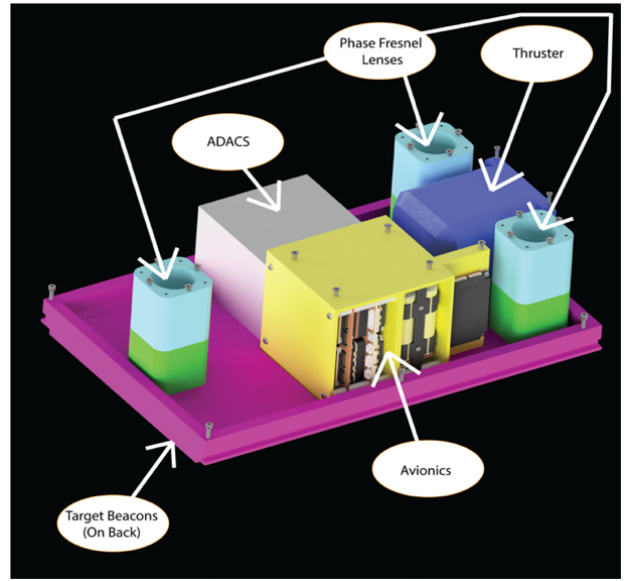

*Figure 12: Diagram of the internal components of the Lens Sat*

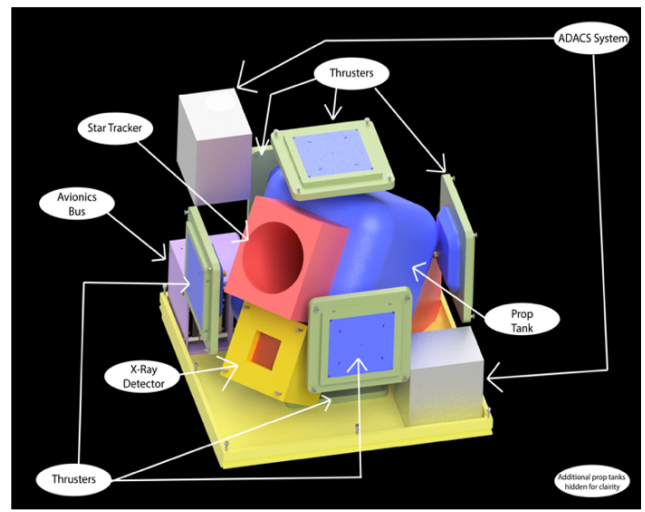

*Figure 11: Internals of the Detector Sat with propellent tanks hidden*

## 12. Conclusion

The VTXO mission is being designed be able to provide X-ray observations at angular resolutions better than a factor of ~10 or better than the current state of the art. It should be able to perform 750 hours of observation over a period of about 6 months. Additionally, further research is being performed on propulsion and trajectory enhancements which may significantly improve these numbers. Finally, the formation flying techniques required for VTXO should be expandable to other applications, including the multi-spacecraft constellations required to build a distributed aperture telescope.